\documentclass[12pt]{iopart}
\pdfoutput=1

\usepackage{iopams}
\usepackage{cite}
\usepackage{url}
\usepackage{color}

\usepackage{graphicx}

\begin{document}
\title[ ]{Heat and work distributions for mixed Gauss-Cauchy process}

\author{{\L}ukasz Ku\'smierz}
\address{Marian Smoluchowski Institute of Physics and Mark Kac Center for Complex Systems Research, Jagellonian University, ul. Reymonta 4, 30--059 Krak\'ow, Poland }

\author{J. Miguel Rubi}
\address{Departament de F\'isica Fonamental, Facultat de F\'isica, Diagonal 647, 08028 Barcelona, Spain}

\author{Ewa Gudowska-Nowak}
 \address{Marian Smoluchowski Institute of Physics, and Mark Kac Center for Complex Systems Research, Jagellonian University, ul. Reymonta 4, 30--059 Krak\'ow, Poland}
\eads{\mailto{lukasz.kusmierz@uj.edu.pl}, \mailto{gudowska@th.if.uj.edu.pl},\mailto{mrubi@ub.edu}}

\date{\today}
 
 \begin{abstract}
We analyze energetics of a non-Gaussian process described by a stochastic differential equation of the Langevin type. The process represents a paradigmatic model of a nonequilibrium system subject to thermal fluctuations and additional external noise, with
 both sources of perturbations  considered as additive and statistically independent forcings. We define thermodynamic quantities for trajectories of the process and analyze 
 contributions to mechanical work and heat. As a working example we consider a particle subjected to a drag force and two  {statistically} independent L\'evy white noises with stability indices
 $\alpha=2$ and $\alpha=1$.  The fluctuations of dissipated energy (heat) and distribution of work performed by the force acting on the system are addressed by examining contributions of Cauchy fluctuations  {($\alpha=1$)} to either bath or external force acting on the system.
 \end{abstract}
\maketitle

\section{Introduction}\label{sec:introduction}

Superlinear and sublinear diffusional transport and nonexponential  relaxation kinetics are ubiquitous in nature and have been observed and analyzed in a number of systems ranging from hydrodynamic flows and plasma physics to transport in crowded environment inside cells and price variations in economy \cite{Klafter}. Usually, a starting point of description of systems exhibiting anomalous transport properties involves use of continuous time random walk (CTRW) and fractional differential diffusion equations of the Fokker-Planck type (FFPE). Although both concepts turned out to be extremely  powerful in applications, there is still lack of full understanding of their links with first principles of mechanics and thermodynamics \cite{Klages,Balescu,Marconi}.

 {The celebrated Smoluchowski-Fokker-Planck equation \cite{Risken} provides a tool to quantify propagation of randomness in stochastic nonlinear dynamical systems and in a standard from has been derived from the stochastic differential equations driven by Gaussian noises. However, by virtue of the generalized central limit theorem put forward by Paul L\'evy \cite{Levy}, the Gaussian distribution is a special case of more fundamental  "stable statistics" which form the class of distributions possesing selfscaling properties under convolution. In particular, summing independent random variables  having this distribution results in a random variable with a similar distribution to the original ones. This attraction property of stable distributions is characterized by the stability index $\alpha$,  $(0<\alpha\le 2)$ which describes the exponent of the power-law of probability distribution tails : $p(\xi)d\xi\approx |\xi|^{-(1+\alpha)}d\xi$. In analogy to the Wiener process representing realization of diffusive Brownian motion, the L\'evy process $\{X(t), t\ge 0\}$ can be then described as a continuous, homogeneous stochastic process with independent increments sampled from the stable L\'evy distribution and converging to the common Gaussian case for $\alpha=2$. Accordingly, formal time derivatives of L\'evy processes can be understood as examples of white (Markov) noises with (in general) non-Gaussian ($\alpha\neq 2$) distributions of jumps. In consequence, similarly to the correspondence between Smoluchowski-Fokker-Planck equation and stochastic differential equations driven by Gaussian noises, derivation of fractional Fokker-Planck equation  \cite{Yanovsky,Klafter} establishes the link between the Langevin-type equation with non-Gaussian stable white noise and temporal evolution of probability distribution of the ensemble particles whose motion it describes. }

 Among various problems addressed in the field  {of L\'evy-noise driven dynamics} is the origin of superdiffusive transport in momentum space investigated in a number of studies \cite{Gonchar,Trigger,EbRoSo09,EbSoGu13}. In particular, based on the Langevin model with linear friction proportional to velocity and non-Gaussian noise,
superdiffusive transport in velocity space for a magnetized plasma has been analyzed \cite{Gonchar}. Resulting FFPE  has been further shown to describe the evolution of the velocity distribution function of strongly nonequilibrium and hot plasmas obtained in tokamaks \cite{EbRoSo09}.

On the other hand, anomalous transport is also relevant for analysis of signal transmission and detection in biological systems \cite{Heermann,Matthaus}. 
In this field Brownian-ratchet models provide mechanisms by which operation of biomolecular motors is investigated \cite{Bier,Ciliberto,EwaFias2008}.
In recent analysis it has been shown  that the minimal setups of simple ratcheting potentials and additive white symmetric L\'evy (non-Gaussian)  noise are sufficient to produce
 directional transport {\cite{Bartek2004,Dyb_EGN2008,Negrete2008,Ilya2010} and inversion of currents can be obtained by considering a time periodic modulation of the chirality of the L\'evy noise \cite{EwaBarIg,Ewa,Bartek}.
 
Here we continue this line of research and analyze thermodynamic energetics of a linear system subject to L\'evy fluctuations.  We assume that the motion of a test particle confined by a harmonic potential is described by a Langevin equation
\begin{equation}
\dot{x}(t) = -a (x(t) - v t) +\sqrt{2}\sigma  \xi_G(t) + \gamma \xi_C(t)
\label{model}
\end{equation}
In the above equation $x$ represents a position at time $t$, $a$ is a parameter related to the strength of the harmonic potential and friction coefficient, and $\xi_G(t)$, $\xi_C(t)$ stand for independent white Gaussian ($\alpha$=2) and Cauchy ($\alpha=1$) stable noises.
Since the dynamics is analyzed in the overdamped limit, the particle has fully equilibrated (Maxwellian distributed) velocities and its internal energy is solely given by the potential energy which includes comoving frame, i.e. $U(x,t)=(x(t)-vt)^2/2$. Models of that type with only Gaussian noise term entering Eq.(\ref{model}) have been previously successfully applied to describe motion of colloidal particles in optical tweezers and unfolding of biopolymers \cite{Ciliberto,Bustamante,Sekimoto,Rubi}. They have been also test arrangements used in asserting a character of fluctuation theorems for measurable quantities and their symmetry.            {For thermostated dynamical systems of that type, interacting with thermal reservoirs, generalization of the fundamental second law of thermodynamics has led to derivation of so called fluctuation theorems (fluctuation relations) \cite{Chechkin}.
The latter express large-deviation symmetry properties} of the probability density functions for physical observables considered in nonequilibrium situations. They have been thoroughly analyzed for systems operating close to equilibrium, when fluctuations of thermodynamic variables follow the Gaussian law of statistics. Extensions to non-Gaussian white noises and correlated Gaussian fluctuations have been also recently addressed \cite{Chechkin,Lenz,Cohen,Touchette}.
 In particular, for linear systems perturbed by L\'evy white noise, it was found that the ratio of the probabilities of positive and negative work fluctuations
of equal magnitude behaves in an anomalous nonlinear way, developing a convergence to one for large fluctuations \cite{Touchette,Cohen,Chechkin}. Hence, negative fluctuations of the work performed on the particle are just as likely to happen as large positive work fluctuations of equal magnitude. This unusual property strongly departs from the Gaussian case, where the validity of a standard FR implies that positive fluctuations are exponentially more probable than negative ones. Our analysis further generalizes those findings and refers to a situation when the source of fluctuations is modeled by two independent white noises: Cauchy noise of intensity $\gamma$ and Gaussian noise of intensity $\sigma$, both entering the system as random additive forces. 

          {The paper is organized as follows: Section 2 provides a brief recapitulation of  thermodynamic interpretation of a standard Langevin equation, in which the source of  random forces is represented by a generalized  derivative of a Gaussian  Wiener process (white noise). In Section 3 we devise analysis of thermodynamic potentials based on the concept of  the ensemble average of the system trajectories, i.e. utilizing the probability density function of a corresponding Fokker-Planck-Smoluchowski equation. The dynamic behavior of the nonequilibrium system is further  investigated in Sections 4 and 5  in terms of fluctuating thermodynamic components of energy like heat, mechanical and dissipated work and internal energy.
Crucial for this analysis is statistical balance of total energy (Section 6) and fluctuation relations for thermodynamic quantities, discussed in Section 7. Section 8 brings about summary of the results. }

 \section{Thermodynamic description: the entropy production in a system subjected to external driving forces}
 Following derivation of de Groot and Mazur \cite{Mazur,Veinstein,Rubi2012}, we first note that the external field ${\mathbf F}$ acting on the system increases its internal energy $U$ by an amount $dU={\mathbf F} \cdot d {\mathbf x}$, where $\mathbf x$ denotes a state variable coupled to the action of the $\mathbf F$ force. The differential of entropy can be expanded in variables $\mathbf x$ and $U$ (both assumed to be independent variables) yielding 
 \begin{equation}
 dS=\left (\frac{ \partial S} {\partial U}\right )_{x}dU+\left (\frac{ \partial S} {\partial {\mathbf x}}\right )_{U}dx = \left ( \frac{{\textbf F}}{T}-g\cdot\mathbf x \right )dx
 \label{entropy}
 \end{equation}
 In the linear regime of nonequilibrium thermodynamics we identify
 stationary states of the system with thermal equilibrium for which probability $p({\bf{x}} )d{\bf{x}}$ that a system is in a state for which $x$ takes on values between $x$ and $x + dx$ can be expressed as $p(x)dx=C\exp[S(x)/k_B]=C \exp[-\frac{1}{2}k_B^{-1}\sum g_{ij}x_jx_i]$. Accordingly,  $g$ stands for the matrix whose elements are $g_{ij}=(\partial^2S/\partial x_i\partial x_j)_U$ and we introduce $\mathbf{X}_i=\frac{\partial S}{\partial x_i}=-g\cdot\mathbf x$, an intensive thermodynamic state variable (generalized thermodynamic force) coupled to an extensive variable $\mathbf x$. For driving forces which are constant in time, the linear response \cite{Foot,EbSoGu13} allows us to establish a relation for the mean value $\left<\mathbf x(t)\right > =\hat{\chi}(0)\mathbf{F}$ which is identified with the most probable value of $\mathbf x$ in the stationary ensemble \cite{Mazur}.
Based on Eq.(\ref{entropy}), the entropy production in the system is described by
 \begin{equation}
 \frac{dS}{dt} = (\frac{\mathbf{F}}{T} + \mathbf{X})\frac{d\mathbf x}{dt}.
 \label{prod}
 \end{equation} 
By use of the the first and second laws of thermodynamics expressed for quasi-stationary processes, the entropy production under constant temperature  can be rewritten as
\begin{eqnarray}
\frac{dS}{dt} = \frac{1}{T}\frac{d}{dt} \left \{dU-DW+pdV-\mu dN \right\}=\nonumber \\
 \frac{1}{T}\frac{d}{dt} \left \{dU-DW \right\}|_{V,N},
\label{entropy_production}
\end{eqnarray} 
where the symbol $DW$ (and in the forthcoming paragraphs, $DQ$) has been used to denote the general Pfaffian differential forms  (to be distinguished  from the exact differential, to which they can be turned into by finding an appropriate integrating factor).
As an exemplary case, we consider further time variations of the variable $x$ described by the Langevin equation for the one-dimensional damped harmonic oscillator of a unit mass:
\begin{equation}
\ddot{x}=-\Gamma\dot{x}-kx+F+\tilde{\xi}_G(t).
\end{equation}
Here $\Gamma$  stands for the friction constant related to the correlation function of Gaussian white noise
\begin{equation}
 \left<\tilde{\xi}_G (t)\tilde{\xi}_G(t') \right >=2\Gamma k_BT \delta (t-t').
 \label{FD}
 \end{equation}
 In accordance with Eq.(\ref{prod}), the thermodynamic force coupled to the coordinate $x$  is of the Hooke's law type $\mathbf{X}=-kx/T$. Examination of time variations of mechanical energy of the system yields:
\begin{equation}
\frac{d}{dt}\left [\frac{1}{2}\dot{x}^2+k\frac{x^2}{2}\right ]=
\left [-\Gamma \dot{x} +F+\tilde{\xi}_G(t)\right ]\frac{dx}{dt}
\label{energy_balance}
\end{equation}
By combining Eqs.(\ref{energy_balance}) and (\ref{entropy_production}), we identify heat transferred from the system to the heat bath as
\begin{equation}
 -DQ\equiv\left (\Gamma\frac{dx}{dt}-\tilde{\xi}_G(t)\right )dx,
 \end{equation}
 and work performed on the system $DW\equiv Fdx$.
 In the overdamped limit, equation of motion for variable $x$ takes on the form
 \begin{equation}
 \dot{x} = -\frac{k}{\Gamma}x+\frac{F}{\Gamma}+\frac{1}{\Gamma}\tilde{\xi}_G(t), 
 \label{overdamped}
 \end{equation}
 so that $1/\Gamma=\tau$ represents time scale of relaxation of velocities to the stationary state.
Accordingly, the energy conservation law can be expressed by multiplying the forces in  Eq.(\ref{overdamped}) by an incremental change of the state  $dx$ and identifying heat discarded by the system into the bath $DQ$ (or heat transferred from the bath into the system, \cite{Sekimoto,Seifert}), internal energy $dU$ and work performed by the external force $dW$:
\begin{eqnarray}
( -\Gamma\dot{x} +\tilde{\xi}_G(t) - kx+F )dx = \nonumber \\
 DQ-dU+dW=0
\end{eqnarray}
In the absence of additional Cauchy random force, Eq. (\ref{overdamped}) transforms ($\gamma=0$) into (\ref{model}) by identifying $a={k}{\Gamma}^{-1}$, ${\tilde{\xi}(t)}{\Gamma}^{-1}=\sqrt{2}\sigma \xi(t)$ and ${F}{\Gamma}^{-1}=avt={k}{\Gamma}^{-1}vt$. Now, the Hooke's force stems from the potential pulled with a constant velocity $v$.
In a comoving frame $y=x-vt$, the model system Eq.(\ref{model}) can be further reduced to
\begin{equation}
\dot{y}(t)=-ay-v+\sqrt{2}\sigma\xi_G(t)+\gamma \xi_C(t)
\label{frame}
\end{equation}
with the coefficient $a$ established by the relation $a=k/\Gamma$.
Consequently, for negligible Cauchy noise contribution ($\gamma=0$), in line with the definition Eq.(\ref{prod}) the instanteneous entropy production can be defined at the trajectory level and written in the form of
\begin{equation}
\dot{S}=-\frac{1}{T}\left [ay (\dot{y}+v\right)],
\label{ent}
\end{equation}
with the heat transferred to the system identified as
\begin{equation}
Q\equiv T \int dt \dot{S}(\dot{y},y)
\label{hea}
\label{production}
\end{equation}
where $T$ stands for the ambient temperature.           {When averaged over ensemble of trajectories (i.e. with respect to the PDF $p(x,t)=<x(t)-x>_{\xi_G(t)}$), Eqs.\ref{ent}, \ref{hea} reflect increase of the entropy in the medium $\dot{S}_m=-\frac{1}{T}<ay(\dot{y}+v)>_{p(x,t)}$, in line with definition of stochastic energetics \cite{Seifert,Sekimoto}. }
It has to be stressed that a problem of study of the action of an additive Gaussian white noise on a Brownian particle, as described by a standard Langevin equation, has  a well documented universal statistical physics approach which, by use of fluctuation-dissipation theorem, relates thermal noise intensity to the friction coefficient of temperature of the system
\cite{Mazur,Sekimoto,Seifert}. This is however,  not the case  of white noises with a heavy tailed distribution of impulse intensities which are qualitatively very different from standard noises characterized by probability distribution densities with finite cumulants (moments) \cite{Chechkin,Lenz,Cohen,Touchette}. In contrast to regular, Gaussian-like noise, the variance of more  general L\'evy  noise is infinite and its power spectral density does not exist. Clearly, this feature violates classical fluctuation-dissipation theorem and obscures thermodynamic interpretation of stochastic Langevin equation by posing dubiety on definition of temperature of the heat bath and its relation to fluctuations imparted to the system \cite{Sekimoto,Mazur,Ewa2012,EbSoGu13}.  Such censure of L\'evy fluctuations is clearly legitimate if the noise is treated as internal (arising from the environment  directly surrounding the test particle). If, on the other hand, the L\'evy noise is considered to be external, time dependent driving force, the friction and external noise are decoupled and the fluctuation-dissipation relation Eq.(\ref{FD}) has to be fulfilled solely by the thermal (internal) part of the noise.    
 
In what follows we will reconsider contributions to entropy production in a linear system driven by independent Gauss and Cauchy noises by analyzing, instead of direct use of the relation Eq.(\ref{production}), the balance of energy with fluctuating work and heat being exchanged in the system.

\section{Irreversible thermodynamics for continuous time stochastic Markov process}
Time evolution of the probability density function (PDF) of the  system described by Eq.(\ref{frame}) is associated with
the fractional FFPE:
\begin{eqnarray}
\label{FFPE}
\frac{\partial p(y,t)}{\partial t}= & & -\frac{\partial}{\partial y}\left[- ay-v \right] p(y,t) + 
  \sigma^2 \frac{\partial^2}{\partial y^2} p(y,t)
 + \gamma \frac{\partial}{\partial |y|}  p(y,t),
\end{eqnarray}
 {which can be derived by use of the cumulant generating  function of the motion defined by the Langevin equation \cite{Yanovsky,Gonchar}.}
Here, the fractional derivative is understood in Riesz-Weyl sense \cite{Yanovsky} and defined by its Fourier transform $\mathcal{F}\left[\frac{\partial^\alpha}{\partial|x|^\alpha}f(y) \right]=-|k|^\alpha \mathcal{F}\left[ f(y)\right]$.   {Consequently,} Eq. (\ref{FFPE}) has the following Fourier representation:
\begin{eqnarray}
\label{eq:fourierrepresentationfp}
\frac{\partial \hat{p}(k,t)}{\partial t} & = &  -ak\frac{\partial}{\partial
k} \hat{p}(k,t) +ikv\hat p(k,t)-
 \sigma^2 |k|^2\hat {p}(k,t)- \gamma|k|\hat {p}(k,t),
\end{eqnarray}
where $\hat{p}(k,t)=\mathcal{F} \left[ p(y,t) \right]$. Note, that in Eq.(\ref{FFPE}) the strength of the white Gaussian noise $\sigma^2$ is related to the temperature of the thermal bath via the usual relation $<\xi_G(t)\xi_G(t')>=2k_BT\Gamma\delta(t-t')=2\sigma^2\delta(t-t')$ with the viscous drag coefficient $\Gamma \equiv 1$ referring to the time units $t=1/\Gamma$.
In contrast, the intensity $\gamma$ of the external fluctuating force $\xi_C$ is an arbitrary parameter of the model.

Due to linearity of the equation (\ref{frame}), PDF of the process attains the form of the convolution of L\'evy distributions (with unknown, so far, time-dependent parameters). The corresponding characteristic function is then:
\begin{equation}
\hat{p}(k,t) = e^{i k \mu (t) - \sigma^2(t) |k|^2 - \gamma (t) |k| }
\end{equation}
Using this ansatz and FFPE (eq. (\ref{FFPE})) we can easily obtain evolution equations for parameters:
\begin{equation}
\dot{\mu}(t)=-a \mu(t) + v
\end{equation}
\begin{equation}
\dot{\gamma}(t)= \gamma - a \gamma(t)
\end{equation}
\begin{equation}
\frac{d \sigma^2(t)}{dt}=\sigma^2 - 2 a \sigma^2(t)
\end{equation}
Accordingly, the stationary solution to the FFPE subject to conditions $\dot{\mu}(t)=\dot{\sigma}^2(t)=\dot{\gamma}(t)=0$) for a constant  $v$ attains  the form \cite{Ewa2012}:
\begin{equation}
\mu_{\infty}:=\lim_{t\rightarrow \infty} \mu (t) = \frac{v}{a}
\end{equation}
\begin{equation}
\gamma_{\infty}:=\lim_{t\rightarrow \infty} \gamma(t) = \frac{\gamma}{a}
\end{equation}
\begin{equation}
\sigma_{\infty}^2:=\lim_{t\rightarrow \infty} \sigma(t) =\frac{\sigma^2}{2 a}
\end{equation}
Stationary PDF can be expressed by a reverse Fourier transform of the
characteristic function:
\begin{eqnarray}
\hat{p}_s(k) = e^{- \sigma^2_{\infty} |k|^2 - \gamma_{\infty} |k| +ik\frac{v}{a}}\nonumber \\
p_s(y)=\frac{1}{2 \pi} \int\limits^{\infty}_{-\infty} \! \mathrm{d}k\mbox{ } \hat{p}_s(k) e^{-i k y}
\end{eqnarray}
Although it cannot be expressed in terms of elementary functions,  for $v=0$ it can be rewritten \cite{EbSoGu13} in a closed form by use of the Faddeeva function $
w(x):=e^{-x^2} \mathrm{erfc}(-i x)$
with $\mathrm{erfc}(x)$ standing for the complementary error function.  Accordingly, stationary PDF solution takes on the  form

\begin{equation}
\label{stationarysolution}
p_s(y)=\frac{1}{2 \sqrt{\pi} \sigma_{\infty}} Re\;\; w(\frac{-y+i \gamma_{\infty} }{2 \sigma_{\infty}})
\end{equation}
          {and assumes a Gibbsian, equilibrium profile with well defined first and second moments only in the absence of the Cauchy-noise contribution, i.e. for $\gamma=0$. Despite that observation,
for the Markovian system described  by Eq.(\ref{frame}), a general version of the H-theorem  has been shown to hold \cite{Vlad}, after identifying the H-functional with the relative entropy. Considering a correspondence with stochastic energetics under Gaussian noise \cite{Seifert,Hatano,Hanggi,Maes}, one is prompted to define a fluctuating entropy function following the Shannon-Gibbs prescription
$S[p(y,t)] = -\int p(y,t)\log{p(y,t)}dy$.  By an analogy to the Gibssian equilibrium state the functional $\Phi(y)\equiv-T\log p_s(y)$ can be then interpreted as a nonequilibrium pseudo-potential \cite{Kubo2,Graham} with 
 equivalents of internal and free energies of the system given by
\begin{equation}
 U=<\Phi(y)>_{p(y,t)}=-T\int p(y,t)\log p_s(y)dy
 \end{equation}
and
\begin{equation}
 F=U-TS=<\Phi(y)>_{p(y,t)}-T<\log p(y,t)>_{p(y,t)},
\end{equation}
respectively. It follows then that all transient solutions to Eq.(\ref{FFPE}) tend towards the stationary nonequilibrium state described by $p_s(x)$. It should be stressed however, that such thermodynamic analogy lacks full integrity with statistical interpretation of thermodynamic quantities and their relation to variations of entropy and precise meaning of the stationary solution.}

          {The interplay of Gaussian and L\'evy noises in the dynamics of the linear system Eq.(\ref{frame}) results in scaling properties of the PDF $p(y,t)=1/C(t)\tilde{p}(y/C(t))$, cf. \cite{EbSoGu13}. Accordingly, the dynamic entropy 
\begin{equation}
 S[p(y,t)] =-\int\tilde{p}(z)\log \tilde{p}(z)dz + \log C(t)\;\;\;\;\;\;
 z=y/C(t)
 \label{Lyap}
 \end{equation}
  increases in time and its production attains the form
  \begin{equation}
\frac{dS}{dt}=\frac{\dot{C}(t)}{C(t)}.
\end{equation}
In particular, for systems driven by  L\'evy white noise described by the stability index $\alpha$, the variation of entropy is governed by a scaling function $C(t)\propto t^{1/\alpha}$, i.e. $S[p(y,t)] =1/\alpha\log t + const$ and accordingly, 
          {
\begin{equation}
\frac{dS}{dt}=(\alpha t)^{-1},
\end{equation}
so that  for increasing $\alpha$ the entropy rate decreases \cite{Prehl}.} This observation has been further  used as a concept in the {\it diffusion entropy analysis}, DEA applied among others,  to characterize statistical properties of natural time series  (as e.g. dynamics of electroencephalogram recordings \cite{Grigolini}) and maximizing information exchange in complex networks \cite{West}. For L\'evy jump diffusion  
 \footnote{L\'evy white fluctuations have been represented in this case by a compound Poisson process $\xi(t)=\sum_m^N \eta_m(t)$ with number of summands $N$  {given by a Poisson distribution and selfsimilar statistics of impulse intensities} $f(\eta)d\eta=const\times|\eta|^{-(1+\alpha)}d\eta$, $2>\alpha>0$.} in a force field, 
Vlad et al. \cite{Vlad} performed 
relaxation
analysis of a Lyapunov function - equivalent to the entropy in Eq.(\ref{Lyap}) - and discusssed existence and stability of the nonequilibrium steady state. In a different context, a general discussion on "measures of irreversibility" and time evolution of relative Shannon and Tsallis entropies constructed for PDFs governed by space-fractional differential operators has been initiated by Prehl et al. \cite{Prehl} who investigated the "entropy production paradox" (i.e. aforementioned decrease of the entropy rate with increasing scaling parameter $\alpha$) for anomalous superdiffusive processes. All those studies point to the significant differences of statistical properties of systems driven by Brownian and L\'evy noises while stressing uncertainty of rigorous thermodynamic interpretation of entropy.}

Accordingly, in the forthcoming sections we do not adhere to a specific formulation of nonequilbrium entropy. As an alternative, we choose direct statistical analysis of fluctuations in work performed on the system and heat transferred to the environment, thus providing the insight into the analogue of entropy production in systems operating close to equilibrium.

\section{Heat and work distribution in the presence of Gaussian noise ($\gamma=0$)}
Let us firstly review heat and work distribution function for a generic Brownian particle system in a confining harmonic potential and in contact with a heat reservoir. Such a scenario is used as a typical model of manipulated nano-systems (e.g. motion of colloidal particles in opticcal tweezers) and biomolecules (unfolding of biopolymers) \cite{Ciliberto,Bustamante,Sekimoto}.  Analytical results can be easily obtained in a steady state regime when the harmonic potential is dragged for a long time ($t_0>>\frac{1}{a}$) before any observables of interest have been measured. In such a case the position of the particle  is a Gaussian random variable with corresponding mean and variance given by
\begin{equation}
\langle y_0\rangle = -\frac{v}{a}
\label{stat1}
\end{equation}
\begin{equation}
\langle y_0^2\rangle -\langle y_0\rangle^2= \frac{\sigma^2}{2a} 
\label{stat2}
\end{equation}
In the absence of dragging ($v=0$) no mechanical work is done on the system. Therefore, dissipated heat is equal to the internal energy change of the particle:
\begin{equation}
Q_t=U_t-U_0=\frac{a}{2}(y_t^2-y_0^2)
\end{equation}
In order to derive the formula for heat PDF we start from the characteristic function:
\begin{equation}
G_{Q_t}(k)=\langle e^{i Q_t k}\rangle = \int\limits^{\infty}_{-\infty} \! \mathrm{d}y_t \! \int\limits^{\infty}_{-\infty} \! \mathrm{d}y_0 \mbox{ } e^{i k \frac{a}{2}(y_t^2-y_0^2)} p(y_t,y_0)
\end{equation}
where $p(y_t,y_0)=p(y_t|y_0)p(y_0)$ is a joint PDF expressing probability of finding the particle at a position $y_0$ at time $t=0$ and finding it at $y_t$ at time $t$. The propagator is given by
\begin{equation}
p(y_t|y_0)=\frac{1}{\sqrt{2 \pi \frac{\sigma^2}{a}(1-e^{-2 a t})}}e^{-\frac{(y_t - y_0 e^{- a t})^2}{2 \frac{\sigma^2}{a}(1-e^{-2 a t})}}
\end{equation}
After integration the closed-form formula for the heat characteristic function reads
\begin{equation}
G_{Q_t}(k)=\frac{1}{\sqrt{1+\sigma^4(1-e^{-2 a t}) k^2}}
\end{equation}
and its inverse Fourier transform yields a final result
\begin{equation}
P_{Q_t}(q)=\frac{1}{\pi \sigma^2 \sqrt{1-e^{-2 a t}}}K_0(\frac{q}{\sigma^2 \sqrt{1-e^{-2 a t}}})
\label{heatPDF}
\end{equation}
where $K_0(x)$ stands for the zeroth order modified Bessel function of the second kind.           {This result is complementary to 
former derivations: in \cite{Zon} the Fourier transform of the distribution of heat fluctuations is obtained for a constant velocity $v$ by observing that $W_t$ is Gaussian distributed, while $Q_t=W_t-U(x,t)$, through definition of $U(x,t)$ is quadratic in trajectory $x$. In turn, works \cite{Fogedby,Imparato} have presented long-time heat PDF under assumption of equlibrium (Boltzmann) distribution of initial positions and $v=0$. Here time-dependent PDF for heat Eq.(\ref{heatPDF}) has been derived by assuming stationary Gaussian form of $p(x_0)$ with parameters Eqs.(\ref{stat1},\ref{stat2}). }

\begin{figure}[!ht]
\begin{center}
\includegraphics[angle=0,width=0.6\textwidth]{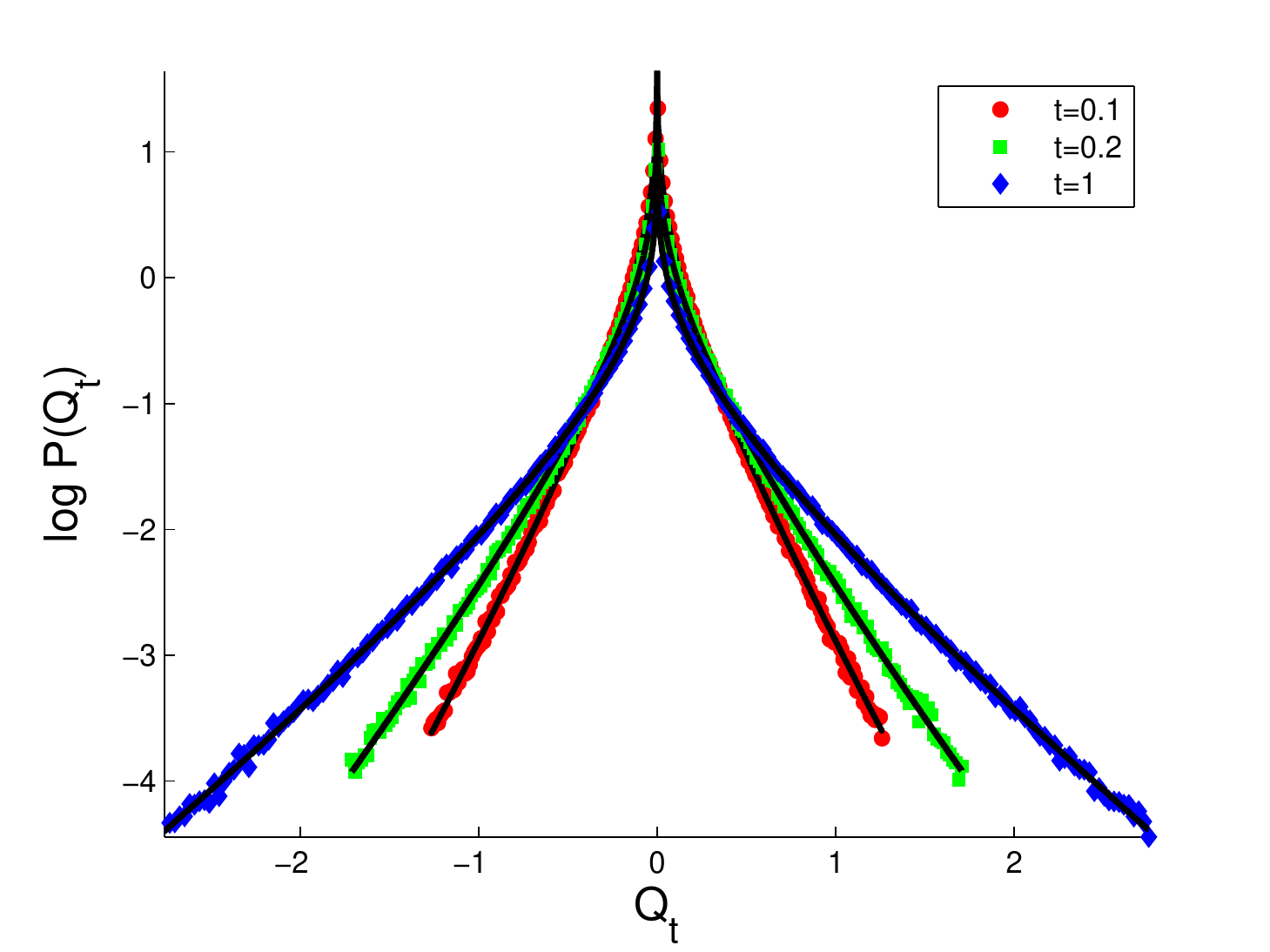}
\caption{Comparison of the heat PDFs obtained with analytical formula,  {Eq.(\ref{heatPDF})} (lines) and numerical estimation (points) for $t=\{0.1, 0.2, 1\}$. Other parameters: $v=0$, $a=1$, $\sigma=1$, $\gamma=0$. Integration step has been set up to $dt=0.1$. Results have been derived for the ensemble of $N=10^6$ sampling trajectories.}
\label{fig:bessel}
\end{center}
\end{figure}

In the case when $v\neq0$ mechanical work performed on the system is given by the formula:
\begin{equation}
W_t=-a v \int\limits^{t}_{0} \! \mathrm{d}s \mbox{ } (x_s - v s)=-a v\int\limits^{t}_{0}ds y_s.
\label{work1}
\end{equation}
Again, due to linearity of the Langevin equation governing evolution of $y(t)$ and the above definition, at any instant of time $W_t$ is a Gaussian random variable with parameters:
\begin{equation}
\langle W_t\rangle = v^2 t
\end{equation}
\begin{equation}
\langle W_t^2\rangle -\langle W_t\rangle^2= \frac{2 v^2 \sigma^2}{a}(a t +e^{-a t} - 1) 
\label{war}
\end{equation}
Figures \ref{fig:bessel} and \ref{fig:VarWt} display congruence between analytical and numerical results, thus validating the numerical algorithm  used in our evaluation of heat and work distributions undrr action of combined Gauss and Cauchy noises. . 
\begin{figure}[!ht]
\begin{center}
\includegraphics[angle=0,scale=0.6]{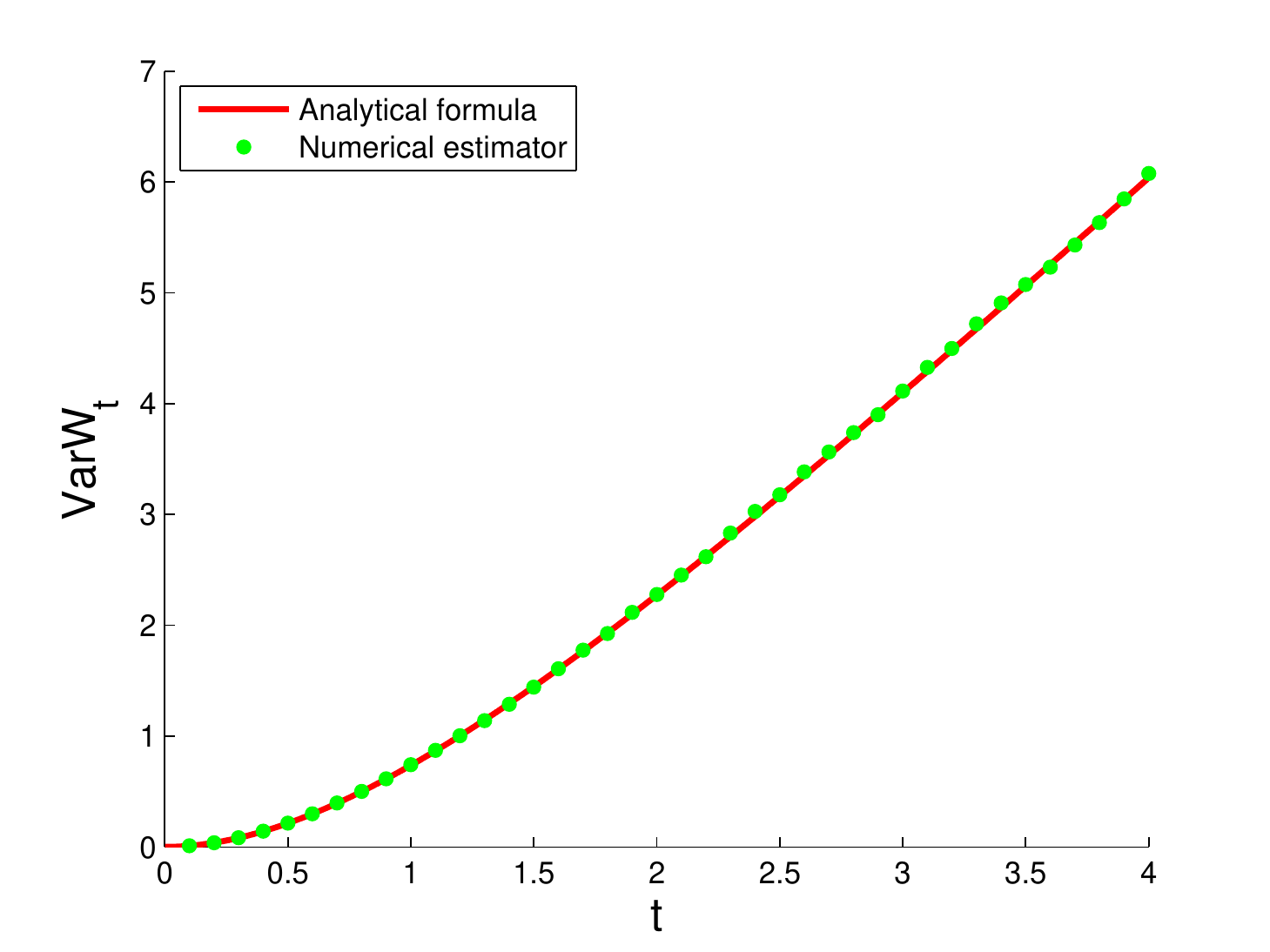}
\caption{Comparison of work variance as a function of time obtained from analytical formula,  {Eq.(\ref{war})} (line) and by numerical estimation (points). Parameters: $v=1$, $a=1$, $\sigma=1$, $\gamma=0$, $dt=0.1$, number of trajectories: $N=10^5$.}
\label{fig:VarWt}
\end{center}
\end{figure}
\section{Analytical solutions for $\gamma \neq 0$}
Mechanical work done by the potential dragging force is still given by Eq.(\ref{work1}).  By using the method of  characteristic functional \cite{Cohen,Touchette,Caceras,Caceras2,Sabha}  of the process $x_t$,  we derive           {(for details, see Appendix)} the characteristic function of work distribution  $G_{W_t}(w)\equiv<\exp(iwW_t)>$:
\begin{equation}
\ln{G_{W_t}}(k)=i v^2 k t - \sigma^2 v^2 k^2(t+\frac{1}{a}(e^{- a t}-1))- \gamma v |k| t
\label{mech_work}
\end{equation}
\label{central}
\begin{figure}[!htbp]
\begin{center}
\includegraphics[angle=0,scale=0.6]{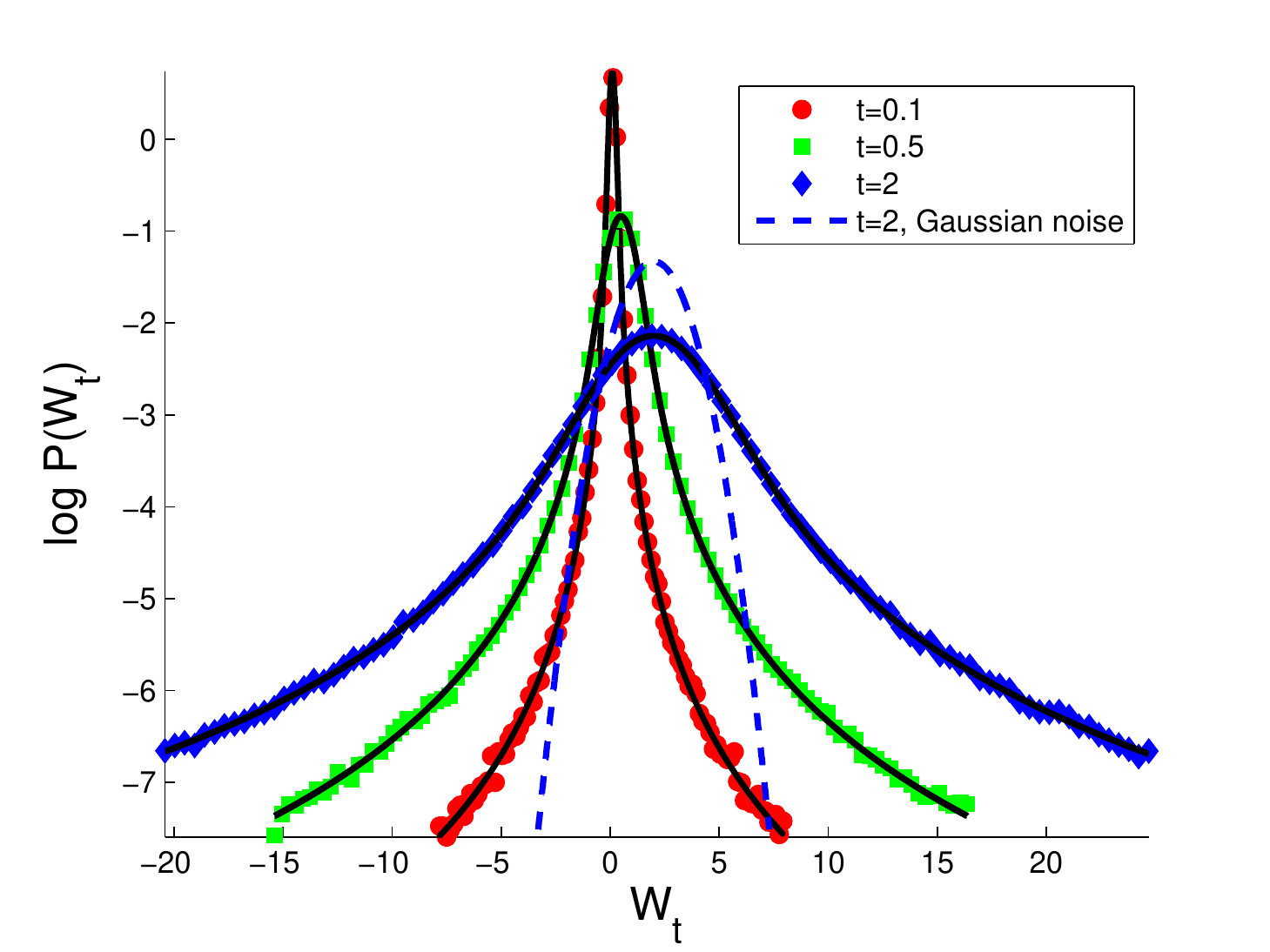}
\caption{Comparison of work PDFs obtained with analytical formula,  {Eq.(\ref{mech_work})} (lines) and numerical estimation (points) at different times $t=\{0.1, 0.5, 2\}$ represented by red, green and blue symbols, respectively. Initial conditions have been sampled from the stationary distribution (in a comoving frame). Seperated dashed line is an analytical result for the Gaussian case ($\gamma=0$ and $t=2$). Other parameters: $v=1$, $a=1$, $\sigma=1$, $\gamma=1$, $dt=0.1$, $N=10^6$.}
\label{fig:Wsteady}
\end{center}
\end{figure}
 
We can also obtain similar result for a non-steady state, e.g. for $x_0$ not fulfilling the requirements Eqs. (\ref{stat1},\ref{stat2}). For initial condition sampled from the delta distribution $p(x,t=0)=\delta(x-x_0)$ one obtains
\begin{eqnarray}
\ln{G_{W_t}}(k) &=& i k [v^2  t - v (x_0+\frac{v}{a}) (1-e^{-a t})] \nonumber \\
&- & \gamma v |k| [t-\frac{1}{a}(1-e^{-a t}) ]\nonumber \\
& -& \sigma^2 v^2 k^2[t+\frac{1}{2 a}(1-e^{-2 a t})\nonumber \\
 &-& \frac{2}{a}(1-e^{- a t})]
 \label{mech_work_non}
\end{eqnarray}
\begin{figure}[!htbp]
\begin{center}
\includegraphics[angle=0,scale=0.6]{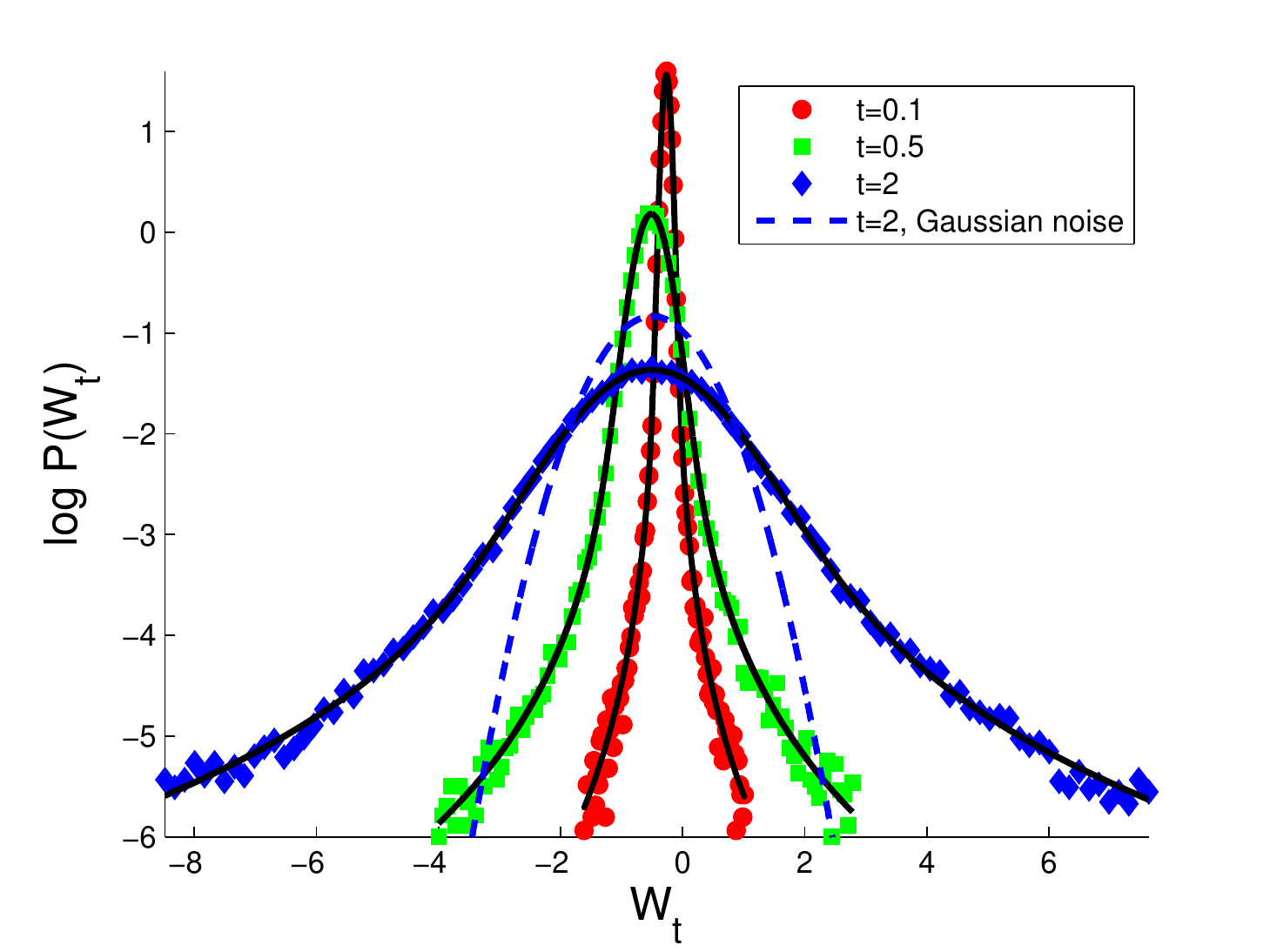}
\caption{Comparison of work PDFs obtained with analytical formula,  {Eq.(\ref{mech_work_non})} (lines) and numerical estimation (points) for $t=\{0.2, 0.5, 1.5\}$ under non-steady state conditions (see the maintext). Seperated dashed line displays an analytical result for Gaussian case ($\gamma=0$ and $t=1.5$). Other parameters: $v=1$, $a=1$, $\sigma=1$, $\gamma=1$, $x_0=\frac{\pi}{2}$, $dt=0.01$, $N=10^5$.}
\label{fig:Wpoint}
\end{center}
\end{figure}
As shown in Figs.\ref{fig:Wsteady},\ref{fig:Wpoint}, the asymptotic form of work PDF $P(W_t)$ is governed by tails of the Cauchy distribution behaving like $|W_t|^{-2}$.
          {Note, that as explained in the following Section, our calculations refer to the situation when additive Gaussian and Cauchy noises acting on the system are assumed
here to be generated by the surroundings and do not contribute to the definition of mechanical work \cite{Ewa,Bartek,EbSoGu13}.}

\section{Numerical schemes used for generating distributions of heat and work}

Unlike Gaussian noise, which is pertinent to situations close to equilibrium, where fluctuations of the components of the Gibbs entropy are governed by Gaussian law, more general infinitely divisible probability laws are known to account accurately for long-range interactions and anomalous fluctuations of physical observables as encountered in strictly non-equilibrium complex systems \cite{Balescu,Garbaczewski,Garbaczewski2}. Typical examples are stochastic fields acting on charged particles in plasma and random graviational forces in clustered systems \cite{Trigger,EbRoSo09,Balescu}, both described by Holtsmark distributions of energy and velocities exhibiting heavy long tails.
In a recent work \cite{EbSoGu13}, we have shown that mixed (convoluted) Gauss-L\'evy distributions are candidates for describing observed deviations from Maxwell distributions in plasmas and other systems \cite{Trigger,EbRoSo09}. They are also proper distributions describing effects of energy transfer in radiation resulting in so called Voigt profile (convolution of Doppler and Lorenz line profiles) of spectral lines.
          {As the addition to those works, in the forthcoming paragraphs we present evaluation of work and heat distributions constrained by the assumption that 1) Cauchy noise is an external driving random force acting on the system, otherwise subject to Gaussian (equilibrated bath) fluctuations and 2)  Cauchy noise is a component of nonequilibrium bath. In both cases our major concern is analysis of the balance of energy and energy exchanged between the system and its reservoir.}

Following definitions of  Section II expressing stochastic energetics \cite{Sekimoto} at the level of trajectories described in terms of  a Langevin equation, we further investigate thermodynamic quantities like work and dissipated heat by exploring their PDFs. That concept can be studied by formulating discretized versions of corresponding stochastic differential equations which we interpret according to Stratonovich.
 In particular, for a system subject to internal Cauchy noise, discretized formulas for heat and work  read:
\begin{itemize}
\item Internal Cauchy noise
\begin{equation} 
\Delta Q_t^{int} = \frac{1}{h}[ \sqrt{2}\sigma w^G_{t,t+h} + \gamma w^C_{t,t+h} - (x_{t+h}-x_t)](x_{t+h}-x_t)
\label{heat}
\end{equation}
where $w^G_{t,t+h}$ and $w^C_{t,t+h}$ stand for (Gaussian- or Cauchy- distributed) increments of the (stationary) Wiener process.
Alternatively, heat exchange can be obtained by direct use of the energy balance formula $\Delta Q=\Delta U-\Delta W$ with the change in internal energy due to the mechanical work given by
\begin{equation} 
\Delta W_t = -k v (x_t - v t) h
\label{praca}
\end{equation}
Both methods of evaluation have been tested for the problem at hand and produced identical results.
In turn, 
for a corresponding situation with external Cauchy-L\'evy noise driving we refer to a discretization scheme:
\item External Cauchy noise
\begin{equation} 
\Delta Q_t^{ext} = \frac{1}{h}[\sqrt{2}\sigma w^G_{t,t+h} - (x_{t+h}-x_t)](x_{t+h}-x_t)
\label{heat_external}
\end{equation}

\begin{equation} 
\Delta W^c_t =  \Delta W_t+\gamma w^C_{t,t+h}\frac{x_{t+h}-x_t}{h}
\label{work}
\end{equation}
\end{itemize}
Note, that in both cases the Cauchy random force does not contribute directly to evaluation of the mechanical work $W_t$.            {When considering $\xi_c(t)$ as an explicit random  force acting on the system, we assess the "total work" PDF by analyzing random quantity $W^c_t=W_t+W^r_t$, i.e. we add extra (random) energy $W_t^r$ from the Cauchy noise to the overall definition of work performed on the system.}

In order to achieve convergence for estimated PDF for heat, a correct integration method  requires that:
\begin{equation}
0=-\frac{d U(x^*_t)}{d x} h - (x_{t+h}-x_t) +w_{t,t+h}
\end{equation}
where $x^*_t = \frac{x_t + x_{t+h}}{2}$, i.e. the Stratonovich rule of integration has to be applied \cite{Sekimoto}.

Moreover, we assume that a proper non-trivial spatio-temporal structure of environmental fluctuations can be preferentially modeled by a smooth colored noise process which  achieves the limit of a somewhat ill-defined idealisation, namely the white noise, as the correlation time of the approximation tends to zero. Indeed, according to  the Wong-Zakai theorem  \cite{Wong,Kupf,Sus,Mannella,Kanazawa} in this limit the smoothed stochastic integral converges to the Stratonovich stochastic integral.
Naively, one could think that e.g. in order to achieve a white noise limit of the Ornstein-Uhlenbeck process, one should scale the corresponding SDE
\begin{eqnarray}
d\xi(t)=-\frac{\xi}{\epsilon^2}dt+\frac{\sigma}{\epsilon}dW(t)\equiv -\Gamma\xi dt=\hat{\sigma}dW(t)
\end{eqnarray}
with the parameter $\epsilon\rightarrow 0$. Such scaling yields formula for the correlation function of the process
\begin{eqnarray}
C(|t-s|)\equiv\left<\xi(t)\xi(s)\right>=\frac{\hat{\sigma}^2}{2\Gamma}\exp(-\Gamma |t-s|)=\nonumber \\
\frac{\sigma^2\epsilon^2}{2}\exp (-\frac{|t-s|}{\epsilon^2})
\end{eqnarray}
with the characteristic correlation time of the noise given by $\Gamma^{-1}$. The spectral density of the rescaled process is then
\begin{eqnarray}
S(\nu)\equiv\frac{1}{2\pi}\int e^{-i\nu \tau}C(\tau) d\tau = \frac{\hat{\sigma}^2}{2\pi(\nu^2+\Gamma^2)}=\frac{\sigma^2\epsilon^2}{2\pi(\nu^2\epsilon^4+1)}
\end{eqnarray}
and tends to $0$ for $\epsilon\rightarrow 0$ which indicates that we end up with the "noiseless" limit. In contrast, the proper white noise limit can be achieved when at the same time
 $\sigma\rightarrow\infty$ and $\Gamma\rightarrow\infty$ in such a way that $\sigma^2/2\Gamma\rightarrow const$. Under these circumstances the power density $S(\nu)$ tends to a constant which is signature of the white noise.
 In other words, by taking e.g. the rescaled version of the Gaussian colored noise $\xi_G(t/\epsilon^2)$ in the Langevin equation with multiplicative noise term
\begin{eqnarray}
\dot{x} = a(x) + \frac{f(x)}{\epsilon}\xi_G(t/\epsilon^2), \nonumber \\
<\xi_G(t)\xi_G(s)=\frac{1}{2}\exp(-\frac{|t-s|}{\epsilon^2}),
\label{WZ1}
\end{eqnarray}
in the limit of $\epsilon\rightarrow 0$ the result of integration  converges weakly to $\hat{x}(t)$ which satisfies SDE
\begin{equation}
d\hat{x}=a(\hat{x})dt+f(\hat{x})\circ dw(t)
\label{WZ2}
\end{equation}
where $w(t)$ stands for a standard 1-dim Wiener process. Since our analysis of (stochastic) work and heat functionals relies on estimation of multiplicative white noise sources entering Eqs.(\ref{heat}-\ref{work}), we adjust to the aforementioned scheme, in which the white-noise limit solutions are achieved as limits of integrals performed with respect to colored noise sources.

\begin{figure}[!htbp]
\begin{center}
\includegraphics[angle=0,scale=0.9]{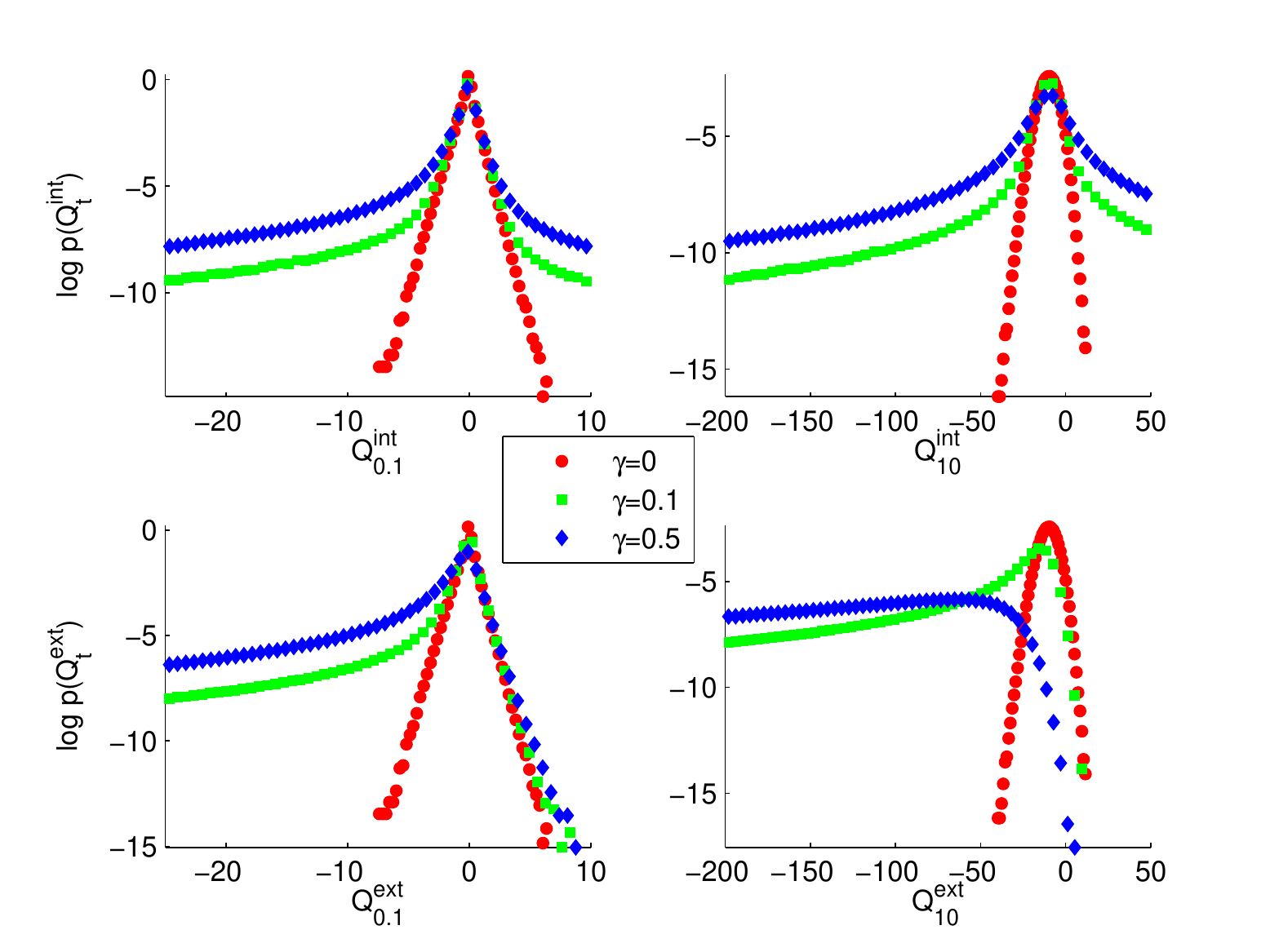}
\caption{          {Short time ($t=0.1$, left panel)  and long time ($t=10$, right panel) heat PDFs for the system subject to mixed Gaussian-Cauchy fluctuations.  Plots represent PDFs for different intensities $\gamma$ of the Cauchy  additive noise. Parameters of the model: $v=1$, $a=1$, $t=0.1$, $\sigma=1$, correlation time of the noise $\epsilon=0.01$. Histograms have been collected on $N=10^6$ trajectories, time step of simulations $\Delta t=10^{-3}$. Upper panels refer to Cauchy noise included in the bath, Eq.(\ref{heat}) whereas lower panels depict heat PDFs for external Cauchy noise, Eq.(\ref{heat_external}) treated as additional random force acting on the system. }  }
\label{fig:Q}
\end{center}
\end{figure}

\begin{figure}[htbp]
\begin{center}
\includegraphics[angle=0,scale=0.9]{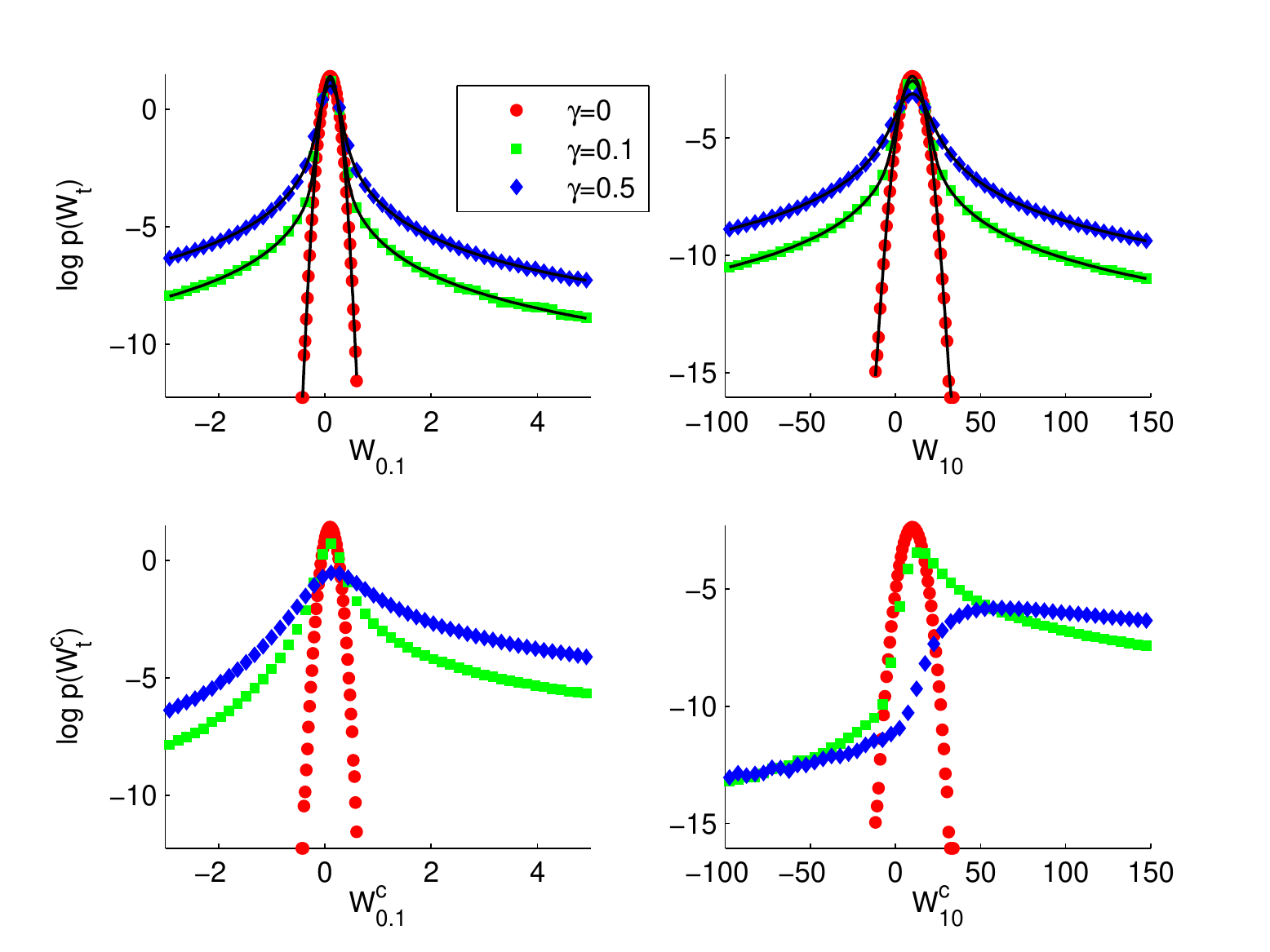}
\caption{          {Short time ($t=0.1$, left panel) and long time ($t=10$, right panel) work PDFs for the system driven by a symmetric internal Gaussian ($\sigma= 1$) noise and an additional Cauchy noise of various intensities $\gamma$.  Parameters of the model: $v=1$, $a=1$, $t=0.1$ correlation time of the noise $\epsilon=0.01$. Histograms have been collected on $N=10^6$ trajectories, time step of simulations $\Delta t=10^{-3}$. Plots in the upper row refer to estimation of the mechanical work Eq.(\ref{praca}) and lower row presents PDFs of the total work under action of steady dragging and additional Cauchy random force, Eq.(\ref{work}). Solid lines indicate analytical results, cf. Eq.(\ref{mech_work}). }}
\label{fig:W}
\end{center}
\end{figure}

{Figures \ref{fig:Q},\ref{fig:W} summarize our findings. First, by using  aforementioned definitions of stochastic heat and work (Eqs. 44-46), the overall balance of energy at the level of a single realization of the stochastic process is acomplished. Second, distribution of energy at the ensemble level changes profoundly, depending on whether the random Cauchy noise is treated as external forcing contributing to the evaluation of the total work performed on the system, or included in the definition of heat. The mechanical part of the energy can be evaluated analytically and numerical results corroborate correctly with formulae derived in the Appendix (cf. upper panel of Fig. (\ref{fig:W})). On  the other hand, estimation of heat or dissipated work  can be achieved either by evaluation of stochastic integrals or by using the energy balance formula of the first law. In both cases, for finite time integration, we obtain almost indistinguishable histograms of derived PDFs.}
\section{Fluctuation relations}
Fluctuations around out-of-equilibrium steady states are frequently described in terms of the probability ratio of a time-integrated observable (like entropy production, heat absorbed by driven Brownian particles or the current of the zero-range processes) evaluated for the forward and reverse changes a system undergoes in course of its evolution.
The ratio has been named fluctuation relation (FR) and by definition serves as a measure of the symmetry in the distribution of  fluctuations \cite{Gallavotti,Kurchan,Zon,Jarzynski}.
In particular, a typical arrangement of a Brownian particle dragged by a spring through a thermal environment modelled by Gaussian white noises has been exploited both experimentally and theoretically to deduce symmetry of FR. It has been shown \cite{Zon,Ciliberto,Sabha} that the work performed on the particle satisfies a standard or conventional FR $P(W_{\tau}=w)/P(W_{\tau}=-w)\approx e^{\rho(w)\tau}$ with $\rho(w)$ being a linear function of $w$. In contrast, 
 the heat fluctuations follow and extended fluctuation relation $P(Q_{\tau}=q)/P(Q_{\tau}=-q)\approx e^{\rho(q)\tau}$
 where the exponential weight that relates the probability of positive and negative fluctuations does not scale linearly with the heat variable, i.e. $\rho(q)$ is  a non-linear function of $q$. As discussed elsewhere \cite{Kurchan,Zon,Touchette,Chechkin}, conventional or extended FRs are valid when the underlying probability distributions obey (for long times $\tau\rightarrow\infty$) a large deviation principle $P(W_{\tau}=w)\approx e^{-\tau k(w)}$ with some rate function $k(w)$. On the other hand, the large deviation principle is violated in self-similar systems, where power law distributions of fluctuations govern the statistics \cite{Chechkin,Touchette}. In such systems the ratio of probabilities of positive and negative work fluctuations of equal magnitude shows anomalous behavior  ($P(W)/P(-W)\approx 1$) exhibiting the same  probability of occurrence  for large positive and large negative fluctuations \cite{Chechkin}.
 
\begin{figure}[!htbp]
\begin{center}
\includegraphics[angle=0,scale=0.9]{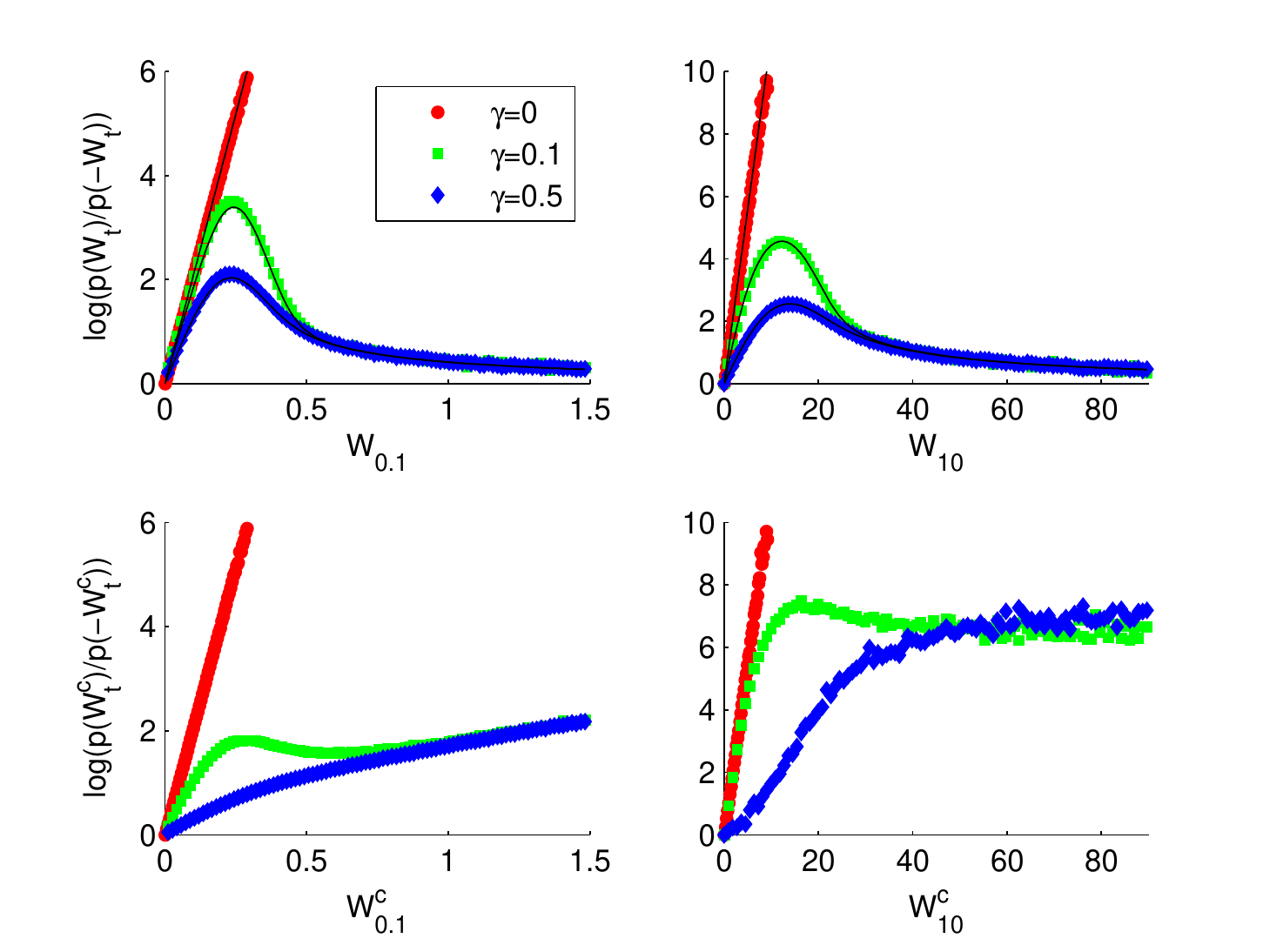}
\caption{          {Work fluctuation (a)symmetry for short time ($t=0.1$, left panel) and long time ($t=10$, right panel). Increments of the mechanical work have been evaluated according to Eqs.(\ref{praca},\ref{work}). Histograms collected on $N=10^6$ trajectories, $a=1$, $v=1$, time step of simulations $\Delta t=10^{-2}$. Solid lines represent analytical results obtained by use of Eq.(\ref{mech_work}). }}
\label{fig:FT}
\end{center}
\end{figure}

Figures (\ref{fig:FT}), (\ref{fig:FTH}) display work and heat fluctuations asymmetry for short and long time limit evaluated for internal Gaussian noise, representing fluctuations of the heat bath and, an external Cauchy noise standing for additional random forcing. In the analysis, the aforementioned Cauchy noise is assumed as either an additional noise in the surroundings or  a random force contributing to the definition of work $\Delta W_t^c$ perforemd on the system, cf. Eq.(\ref{work}). 

\begin{figure}[!htbp]
\begin{center}
\includegraphics[angle=0,scale=0.9]{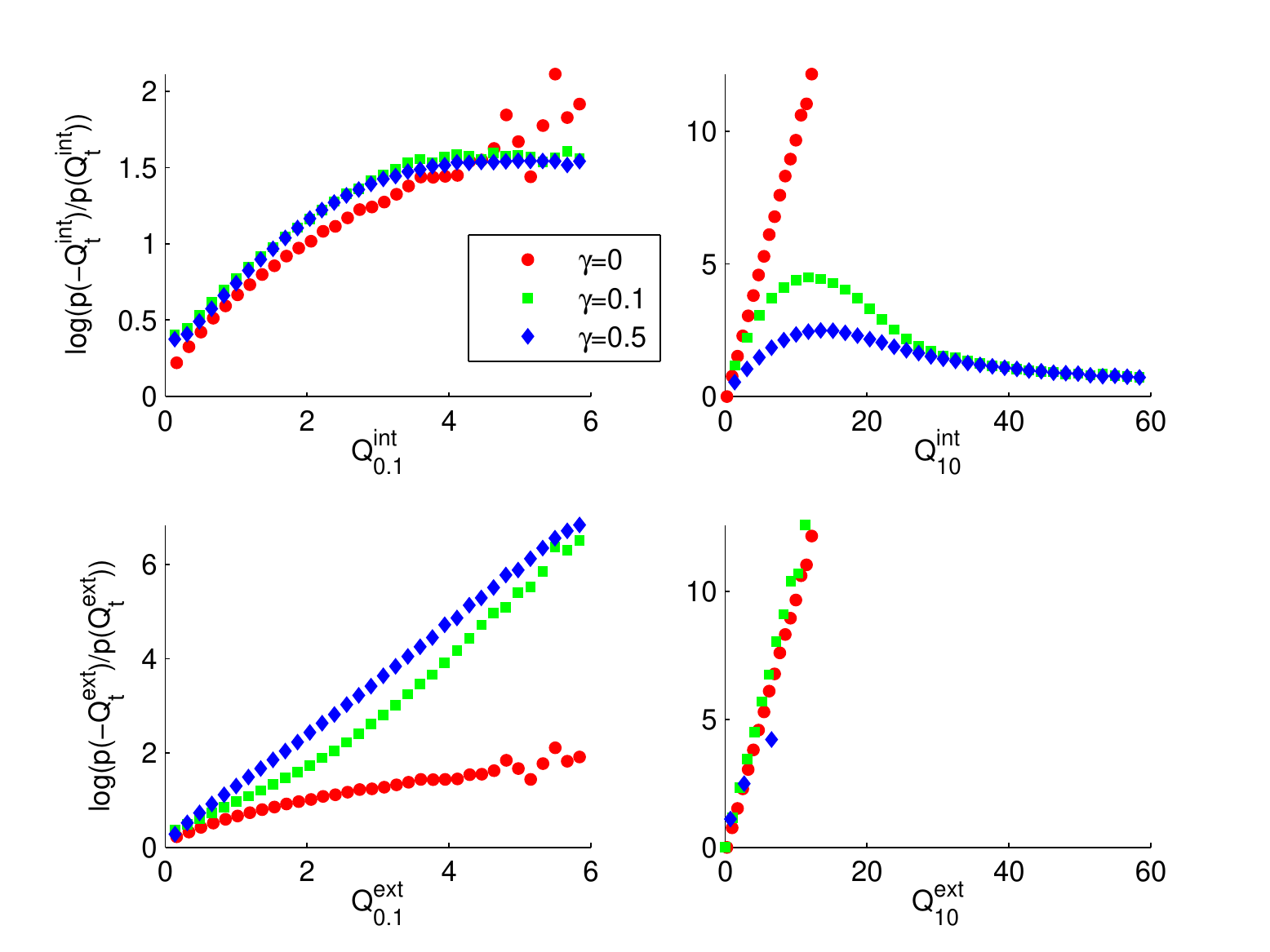}
\caption{          {Heat fluctuation (a)symmetry for short time ($t=0.1$, left panel) and long time ($t=10$, right panel).  Histograms collected on $N=10^6$ trajectories, $a=1$, $v=1$, time step of simulations $\Delta t=10^{-2}$. }}
\label{fig:FTH}
\end{center}
\end{figure}

For a linear system driven solely by Gaussian fluctuations represented by an additive white noise, conventional fluctuation relation is detected, in line with the large deviation probability \cite{Cohen,Sabha,Zon}. In contrast, in the case when the system becomes perturbed by additional Cauchy noise, FR deviates from the symmetric form $P(W_{\tau})/P(-W_{\tau})\approx e^{A\tau w}$ and the assymetry increases with the strength of the Cauchy component in external forcing. For stronger  L\'evy noise, corresponding PDFs do not decay exponentially (reflecting loss of the large deviation principle) and work fluctuation relation assumes the anomalous limit $P(W_t)/P(-W_t)\approx 1$. The location of the maximum of the $P(W)/P(-W)$  ratio depends on the intensity of the Cauchy noise and time of observation:  the assymetry is linear in $W$, it deviates from linearity for increasing values of $W$ and then saturates to a constant value. For longer times
($t=10$), the turnover of the  $P(W_t)/P(-W_t)$  ratio is observed at higher values of $W$ and the slope of the linear consitutent in the ratio is bigger than for short times. Altogether, these observation stay in compliance with the character of the PDFs of increments $\Delta W$ where domination of very large fluctuations leads to the aforementioned asymmetry.

          {Inclusion of the Cauchy white noise into definition of work performed on the system Eq.(\ref{work}) results in heavy tailed, asymmetric PDF of work (see Fig.(\ref{fig:W}) ) resembling mirror-image of heat PDF for system subject to mixed Gaussian-Cauchy fluctuations in the bath (cf. Fig.(\ref{fig:Q})). Also in this case, the fluctuation ratio $P(W)/P(-W)$  reflects the generic asymmetry of work PDF with violation of the exponential large deviation form and recovery of the transient fluctuation relation $P(W_{\tau})/P(-W_{\tau})\approx e^{A\tau w}$ for $\gamma=0$.}

\section{Summary}
We have examined stochastic linear systems subject to action  of  two independent stable noises, one of which exhibits a power law decay of large deviations. In a comoving frame our system represents a generalization of the Ornstein-Uhlenbeck process with  convoluted Gaussian-Cauchy fluctuations. Following standard interpretations of the thermodynamics at the level of the Langevin equation, we have derived formulae for heat and work distribution in such system for various interpretations of the Cauchy noise which enters dynamics either as a contribution to (nonequilibrated) thermal bath, or as an additional external random driving. In order to conform Stratonovich rules of integration, for external Cauchy forces we have adopted white-noise limit Eqs.(\ref{WZ1},\ref{WZ2}) of a colored noise. For Gaussian additive noises, a standard fluctuation relation implies that positive fluctuations are exponentially more probable than negative ones \cite{Cohen,Klages}. This scenario becomes severely affected by the presence of white but otherwise non-Gaussian noises. 
The interplay of the power-law distributed L\'evy fluctuations causes a strong assymetry of the heat PDF and, in line with previous studies \cite{Chechkin,Klages}  leads to anomalous structure of the $P(W_t)/P(-W_t)$  ratio.

Infinite  power of Cauchy noise spectrum results in a breakdown of  the  standard fluctuation-dissipation relationship (FDR) between friction force and correlation function of fluctuations. Nevertheless, when an appropriate conjugate (dynamic) variable is chosen \cite{Ewa2012} in the form of a  derivative of the stochastic entropy in the stationary state $X_c=-T\frac{\partial \Phi(x)}{\partial f}|_{f=0}$ with respect to external driving $f(t)$,  the proper form of FDR can be recasted for sufficiently weak operating noises. This issue, analyzed for the linear system perturbed by Gaussian and Cauchy noises will be the object of our forthcoming studies.


\begin{ack}
The authors acknowledge the support by the European Science Foundation (EFS) through Exploring Physics of Small Devices (EPSD) program.
\end{ack}

\section*{Appendix}

          {Let us recall (cf. Eq.(\ref{frame})) that in the model
 $w(t)=\int\limits_0^t d t' \xi_G(t')$ stands for a standard Brownian motion (Wiener process) and $w_{\alpha}(t)=\int\limits_0^t d t' \xi_C(t')$ is a L\'evy $\alpha$-stable process with the stability index $\alpha=1$. We assume that increments of both processes are statistically independent.}
Our aim is to calculate the characteristic function $G_{W_t}(k)=\langle e^{i k W_t}\rangle$ of the process:
\begin{equation}
W_t=-a v\int\limits_0^t d s (X_{s} - v s) = -a v\int\limits_0^t d s y_s
\end{equation}

We start with the observation that the characteristic functional of the total noise \cite{Caceras,Caceras2,Vlad}  entering Eq.(\ref{frame}) is given by the relation:
\begin{eqnarray}
G_{\xi}[k]\equiv G_{\sqrt{2}\sigma \xi_G+\gamma \xi_C}[k]=G_{\sqrt{2}\sigma \xi_G}[k]G_{\gamma\xi_C}[k] = \nonumber \\
e^{-\sigma^2\int\limits_0^{\infty}d t |k(t)|^2-\gamma \int\limits_0^{\infty}d t |k(t)|}
\label{Gxi}
\end{eqnarray}
Next, we substitute $\xi=\dot{y_t}+a y_t +v$ in the definition of the characteristic function and  integrate the expression with $\dot{y_t}$ by parts:
\begin{eqnarray}
G_{\xi}[k]=\langle e^{i \int\limits_0^{\infty}d t k(t)(\dot{y_t}+a y_t +v) }\rangle=\nonumber \\
\left\langle e^{i(k_{\infty}y_{\infty}-k_0 y_{0})+i\int\limits_0^{\infty}d t[ v k(t) +(a k(t)-\dot{k}(t))y_t]}\right\rangle
\end{eqnarray}
Assuming that $y_0\equiv y_0$ is a number (i.e. $p_{y_0}(y)=\delta(y-y_0)$) and that the function $k(t)$ decays sufficiently fast (securing $k(\infty)=0$), we arrive at:
\begin{equation}
G_{\xi}[k]=e^{-i k_0 y_0+i v \int\limits_0^{\infty}d t k(t)} G_{y|y_0}[-\dot{k}+a k]
\label{GksiodGz}
\end{equation}
In the next step we introduce a new function $m(t)\equiv -\dot{k}(t)+a k(t)$. It is easy to reverse this relation yielding:
\begin{eqnarray}
k(t)=k(0)e^{a t} - e^{a t} \int\limits_0^t d s e^{-a s} m(s)=\nonumber \\
e^{a t} \int\limits_t^{\infty} d s e^{-a s} m(s)
\end{eqnarray}
with $k(0)=\int\limits_0^{\infty} d s e^{-a s} m(s)$. This identyfication together with Eq.(\ref{GksiodGz}) allows us to express the $G_{y|y_0}[m]$  in terms of the known noise characteristic functional:
\begin{eqnarray}
G_{y|y_0}[m]=&e^{i \int\limits_0^{\infty} d s e^{-a s} m(s) z_0-i v \int\limits_0^{\infty}d t e^{a t} \int\limits_t^{\infty} d s e^{-a s} m(s)}\times\nonumber \\
 &\times G_{\xi}[e^{a t} \int\limits_t^{\infty} d s e^{-a s} m(s)].
\end{eqnarray}
Accordingly,
\begin{eqnarray}
\ln{G_{y|y_0}[m]}=&i \int\limits_0^{\infty} d s e^{-a s} m(s) y_0-i v \int\limits_0^{\infty}d t e^{a t} \int\limits_t^{\infty} d s e^{-a s} m(s)\nonumber \\
&- \sigma^2 \int\limits_0^{\infty}d t |e^{a t} \int\limits_t^{\infty}d s e^{- a s} m(s) |^2\nonumber \\
&- \gamma \int\limits_0^{\infty}d t |e^{a t} \int\limits_t^{\infty}d s e^{- a s} m(s) |
\label{GZtfinal}
\end{eqnarray}
Work characteristic functional $G_{W_{t}|y_0}(k)$  can be now obtained by choosing a proper test function $m(t')$: 
\begin{equation}
m(t')=-a v q \Theta(t-t')
\label{properm}
\end{equation}
where $\Theta(\cdot)$ denotes the Heaviside step function. Indeed, by plugging it into formula $G_y[k]\equiv\langle \exp({i \int\limits_0^{\infty}d sk(s)y_s})\rangle$ one can easily verify that this choice leads to $G_{W_t}(k)$.
\newline
The very last step is to calculate integrals in Eq.(\ref{GZtfinal}) with $m(\cdot)$ given by Eq.(\ref{properm}) which leads to the work characteristic function for the dynamics Eq.(\ref{frame}) with an initial condition  $y_0=x_0$:
\begin{eqnarray}
\ln{G_{W_t|x_0}}(k)=&i k [v^2  t - v (x_0+\frac{v}{a}) (1-e^{-a t})] \nonumber \\
&-\gamma v |k| [t-\frac{1}{a}(1-e^{-a t}) ]\nonumber \\
&-\sigma^2 v^2 k^2[t+\frac{1}{2 a}(1-e^{-2 a t})\nonumber \\
&- \frac{2}{a}(1-e^{- a t})]
\label{pierwszy}
\end{eqnarray}
In order to get the $G_{W_t}(k)$ in a steady state regime we have to average  $G_{W_t|x_0}(k)$ over steady state initial conditions at time $t=0$. It is worthy noticing that $G_{W_t|x_0}(k)$ can be rewritten as
\begin{equation}
G_{W_t|x_0}(k)=g(q) e^{-i x_0 b}
\end{equation}
where $b=v q (1-e^{-a t})$ and $g(q)$ does not depend on $x_0$. Thanks to that feature  the final formula contains just the Fourier transform of the initial (steady state) PDF :
\begin{eqnarray}
G_{W_t}(k) =\int d x G_{W_t|x}(k) p_s(x)=\nonumber \\
 g(q) \int d x e^{-i x b} p_s(x)= g(q) \tilde{p}_s(-b),
\end{eqnarray}
and
\begin{equation}
\tilde{p}_s(s)=e^{-i \frac{v}{a} s -\frac{\gamma}{a}|s| - \frac{\sigma^2}{2 a}s^2}
\label{ostatni}
\end{equation}
With the help of Eqs. (\ref{pierwszy}-\ref{ostatni}) we arrive at the final result, Eq.(40t)
\begin{eqnarray}
\ln{G_{W_t}(k)} =
i v^2 k t - \gamma v |k| t\nonumber \\
 - \frac{\sigma^2 v^2}{a} k^2 (a t+e^{-a t }-1).
\end{eqnarray}

\bibliography{citations}
\bibliographystyle{unsrt}
\end{document}